\documentclass[aps,prl,twocolumn,a4paper,10pt,notitlepage,footinbib,superscriptaddress]{revtex4-2}
\usepackage[english]{babel}
\usepackage{lmodern}
\usepackage[utf8]{inputenc}
\usepackage{endnotes}
\usepackage{amssymb,amsmath,amsfonts}
\usepackage{textcomp}

\usepackage{graphicx}
\usepackage{dcolumn}
\usepackage{bm}
\usepackage{xcolor}

\newcommand{\beq}{\begin{equation}}
\newcommand{\eeq}{\end{equation}}
\newcommand{\beqa}{\begin{eqnarray}}
\newcommand{\eeqa}{\end{eqnarray}}

\begin{document}

\title{Schr\"{o}dinger-Navier-Stokes equation for capillary fluids}

\author{L. Salasnich}
\affiliation{Dipartimento di Fisica e Astronomia ``Galileo Galilei''
  and Padua QTech Center,
Universit\`{a} di Padova, Via Marzolo 8, 35131 Padova, Italy}
\affiliation{INFN Sezione di Padova, via Marzolo 8, 35131 Padova, Italy}
\affiliation{corresponding author: luca.salasnich@unipd.it}

\author{S. B\"ohly}
\affiliation{Dipartimento di Fisica e Astronomia ``Galileo Galilei'',
Universit\`{a} di Padova, Via Marzolo 8, 35131 Padova, Italy}

\author{S. Succi}
\affiliation{Center for Life Nano Science@La Sapienza, Istituto Italiano
  di Tecnologia,
00161 Roma, Italy}

\author{A. Tiribocchi}
\affiliation{Istituto per le Applicazioni del Calcolo, CNR,
Via dei Taurini 19, Rome, 00185, Italy}
\affiliation{INFN Sezione di Roma Tor Vergata,
Via della Ricerca Scientifica 1, 00133 Roma, Italy}

\date{\today}

\begin{abstract}
We highlight some properties of the Schr\"{o}dinger-Navier-Stokes (SNS)
equation [Salasnich, Succi, and Tiribocchi (2024)] of potential
relevance for microfluidics and soft matter.
Specifically, we show that the SNS equation with generic parameters
is formally equivalent to the Navier-Stokes-Korteweg equations for
capillary fluids, with the equivalence established at the level of an
action functional that decomposes naturally into a Korteweg conservative
and a dissipative contribution.
We derive the dispersion relation for sound modes, showing that the
dispersive parameter controls capillary stiffness while the dissipative
parameter controls viscous damping, and that the Bogoliubov dispersion
relation is recovered in the quantum limit.
We also derive an effective one-dimensional SNS equation for a fluid
confined in a narrow capillary tube.
\end{abstract}

\maketitle

\section{Introduction}

The study of fluid dynamics through the lens of quantum mechanics has
a long history, dating back to the pioneering work of Madelung
\cite{madelung1926a,madelung1926b}, who showed that the linear 
Schr\"{o}dinger equation for a free particle is equivalent to
the equations of a compressible, inviscid, irrotational fluid at
zero temperature and with zero pressure. This analogy has been extended and
generalized in many directions over the decades
\cite{vauterin1985,coste1998,sala2008,sala2009,molina2016,chavanis2017,yang2023,yang2024},
leading to the recent derivation \cite{sala2024} of a nonlinear 
Schr\"{o}dinger-Navier-Stokes 
(SNS) equation that maps onto the Navier-Stokes
equations for a viscous, irrotational fluid with a generic pressure
at finite temperature. The key ingredient is a shift of the nonlinear potential
of the nonlinear Schr\"odinger equation that 
removes the quantum pressure and introduces a viscous term, while
preserving the conservation of mass. This reformulation naturally connects
quantum pressure effects with Korteweg-type capillarity in classical
fluid dynamics.

The SNS equation contains two dimensionless parameters: $\kappa$,
which controls the transition from the quantum ($\kappa = 0$,
Gross-Pitaevskii-like) to the classical ($\kappa = 1$, Navier-Stokes)
regime, and $\gamma$, which controls the viscous dissipation. The
standard Navier-Stokes equations are recovered for $\kappa = 1$ and
$\gamma \neq 0$, while a Gross-Pitaevskii-like equation is recovered for
$\kappa = 0$ and $\gamma = 0$. Intermediate values of $\kappa$ and
$\gamma$ describe fluids that interpolate between these two limits
and, as we show in this paper, this intermediate regime has a precise
physical interpretation in terms of capillary fluids.

The paper is organized as follows.
In Sec.~II we introduce the SNS equation and its Madelung
representation. In Sec.~III we establish the connection with the
Navier-Stokes-Korteweg equations for capillary fluids.
In Sec.~IV we derive the dispersion relation for
sound modes with generic $\kappa$ and $\gamma$. In Sec.~V we derive
an effective one-dimensional SNS equation for a fluid confined in a
narrow capillary tube.

\section{The Schr\"{o}dinger-Navier-Stokes equation}

Inspired by previous
works \cite{vauterin1985,coste1998,sala2008,sala2009,molina2016,yang2023}, 
in a recent paper \cite{sala2024} we derived a Schr\"{o}dinger-Navier-Stokes 
equation by shifting the nonlinear potential of the Gross-Pitaevskii equation.
The central result is a mapping between a nonlinear Schr\"{o}dinger
equation for a complex field $\psi({\bf r},t)$ and the incompressible
Navier-Stokes equations for a viscous, irrotational fluid.
The mapping is exact and is based on the generalized Madelung transformation
\beq
\psi(\bm{r},t) = \sqrt{n(\bm{r},t)}\, e^{i\theta(\bm{r},t)}
\label{mad1}
\eeq
with fluid number density $n(\bm{r},t) = |\psi(\bm{r},t)|^2$
and velocity
\beq
\bm{v} = D\bm{\nabla} \theta \, ,
\label{mad2}
\eeq
where
\beq
D = {\hbar\over m}
\eeq
is the quantum diffusivity with $\hbar$ the reduced Planck constant and
$m$ the mass of a particle of the fluid. 
The key structural limitation of that formulation is that, since the
fluid velocity is defined as the gradient of the phase field,
the flow is necessarily irrotational, wherever the phase $\theta$ is regular.
Vorticity can only appear in the form of quantized vortices
at phase singularities (points where $\psi = 0$),
which carry circulation in integer multiples of $D$.
However, as expected, it has been recently proved \cite{triola2026} that
a coarse-graining procedure applied to a fully quantum fluid gives rise to an
effectively rotational classical fluid. The SNS equation of Ref.~\cite{sala2024} reads
\beqa
i \hbar \partial_t \psi =  \left[-\frac{\hbar^2}{2m}
  \nabla^2 + \mu(|\psi|^2) + \kappa \frac{\hbar^2}{2m}\frac{\nabla^2
|\psi|}{|\psi|}\right.\nonumber\\\left. +i\gamma \, \frac{\hbar^2}{m}
\nabla^2 \ln\left(\frac{\psi}{|\psi|}\right) \right] \, \psi \; ,
\label{eq:SNS}
\eeqa
where $\mu(n)$ is the bulk chemical potential of the fluid,  such that
\beq
\mu(n) = {\partial \over \partial n} f(n)
\eeq
with $f(n)$ the free energy density, the parameter $\kappa$ controls the
transition from quantum ($\kappa = 0$) to classical ($\kappa = 1$)
regimes, and $\gamma$ is the dimensionless dissipative coefficient.
The standard Madelung picture is recovered in the limit $\kappa = \gamma = 0$.
Note that $\nabla^2\ln(\psi/|\psi|) = i\nabla^2\theta$, so the last
term reduces to $-\gamma(\hbar^2/m)\nabla^2\theta$.

It is important to recall the physical meaning of the $\kappa$ term.
The term $\kappa(\hbar^2/2m)\nabla^2|\psi|/|\psi|$ is precisely the
so-called Bohm quantum potential (more precisely a quantum pressure
term) with sign chosen (for $\kappa=1$) to cancel the quantum
pressure that arises from the kinetic term $-(\hbar^2/2m)\nabla^2$. Indeed, 
for $\kappa = 1$ the quantum pressure is completely cancelled and one
recovers the purely classical NS equations without any density gradient
effects. For $\kappa = 0$ the quantum pressure is fully present and one
recovers the Gross-Pitaevskii equation. For $0 < \kappa < 1$ the
quantum pressure is only partially cancelled, leaving a residual
density-gradient term -- which, as we show below, can be identified with the
Korteweg capillary stress.
We stress that the quantum pressure term is positive definite at the level
of the energy functional and plays the role of a stabilizing interfacial
energy. The terminology ``negative surface tension'' sometimes used in the
literature refers instead to the effective dispersive character of the
corresponding stress term in the dynamical equations, not to a negative
sign of the underlying free energy contribution.

It is worth noting that microfluidics and the SNS equation share a common
mathematical structure: both deal with incompressible flows at low Reynolds
number, where viscous effects dominate and capillary forces at fluid
interfaces play a crucial role. This structural affinity motivates
the present investigation. 
The purpose of this paper is to discuss consequences of
Eq.~(\ref{eq:SNS}) that may be directly relevant to microfluidics.
The first is a formal equivalence between the SNS equation with generic
$\kappa$ and the Navier-Stokes-Korteweg equations for capillary fluids,
established at the level of an action functional. The second is the
derivation of the dispersion relation for sound modes with generic
$\kappa$ and $\gamma$.

\section{Connection with Korteweg Fluids}

\subsection{The Navier-Stokes-Korteweg equations}

By inserting Eqs. (\ref{mad1}) and (\ref{mad2}) into Eq. (\ref{eq:SNS}),
and separating real and imaginary parts, one obtains two equations
for the number density $n({\bf r},t)$ and
the velocity field $\bm{v}(\bm{r},t)$. One is the continuity equation
\beq
    \partial_t n + {\bm\nabla} \cdot (n \bm{v}) = 0 \; .
    \label{eq:continuity}
\eeq
The other is the momentum equation 
\beq
(\partial_t + \bm{v}\cdot\bm{\nabla})\bm{v} =
-\frac{\bm{\nabla} P}{mn} + \nu\nabla^2\bm{v}
+ 
\epsilon \bm{\nabla}\left(\frac{\nabla^2 n}{n}
- \frac{|\bm{\nabla} n|^2}{2n^2}\right)
\; ,
\label{eq:NSK}
\eeq
where $P(n)=n \mu(n) - f(n)$ is the pressure,
\beq
\nu = \gamma \, D = \gamma \, {\hbar\over m}
\eeq
is the kinematic viscosity and
\beq
\epsilon = \frac{{1-\kappa}}{4}\,D^2 
\label{eq:capillary}
\eeq
is the capillary coefficient.
Eqs. (\ref{eq:continuity}) and (\ref{eq:NSK}) are called 
Navier-Stokes-Korteweg (NSK) equations \cite{korteweg1901,dunn1985}. They describe classical fluids
in which the free energy density depends not only on the local density
but also on its gradients. This accounts for capillary effects and
surface tension at fluid interfaces. 
This capillary coefficient $\epsilon$ vanishes
for $\kappa = 1$ (purely classical NS
fluid) and reaches its maximum $\epsilon = D^2/4$ for $\kappa = 0$
(Gross-Pitaevskii-like equation). Intermediate values $0 < \kappa < 1$
describe classical capillary fluids in the NSK sense.

\subsection{Action functional and dissipation}

The last term in Eq. (\ref{eq:NSK}) is the
Korteweg stress and it is responsible for surface tension at fluid interfaces.
It derives from the Korteweg free energy
functional \cite{korteweg1901,dunn1985} 
\beq
\mathcal{F}_K = \int d^3r\left[f(n)
+ \frac{\epsilon}{2n}|\bm{\nabla} n|^2\right] . 
\label{eq:FK}
\eeq
The SNS equation, as we show below, is
associated precisely with the Korteweg functional Eq.~(\ref{eq:FK}),
that is closely related to the Cahn-Hilliard functional \cite{cahn1958}.

Within a generalized Lagrange-d'Alembert-Rayleigh framework for dissipative systems,
the SNS equation can be written as the variational condition:
\beq
\delta\mathcal{S} + 
\delta\mathcal{S}_{\gamma} = 0 ,
\eeq
where 
\beqa
\mathcal{S} = \int dt\,d^3r\left[-\hbar \, n\, \partial_t\theta
  - \frac{\hbar^2 n}{2m}|\bm{\nabla}\theta|^2 - n \, U\right.\nonumber\\\left.
  - (1-\kappa)\frac{\hbar^2}{8mn}|\bm{\nabla} n|^2
- f(n) \right] 
\label{eq:action}
\eeqa
is the conservative action functional in terms of 
number density $n$ and phase 
$\theta$ of the Schr\"odinger 
field $\psi$ with $\delta S$ its first variation, while 
\beq
\delta \mathcal{S}_{\gamma} = \int dt \, d^3r\; \left( - 
\gamma \, \frac{\hbar^2}{m} \,
|\bm{\nabla}\theta|^2 \right) 
\, \delta n    
\label{eq:rayleigh}
\eeq
is the first variation of dissipation. In (\ref{eq:action}) the first term is the Berry phase
(kinetic term for the phase field), the second is the kinetic energy
of the flow, the third is the energy density associated with the external potential $U$, the forth is the residual quantum pressure (proportional
to $1-\kappa$), and the fifth is the bulk interaction energy.
The first variation $\delta \mathcal{S}_{\gamma}$ is the  non-exact 1-form for dissipation, that is strictly related to the dissipative Reyleigh energy functional 
\beq 
{\cal R} = \int d^3r \, \gamma \, {\hbar^2\over m} |\bm{\nabla}\theta|^2 \, n \: . 
\eeq

For a classical fluid, $D$ should be interpreted not as the quantum
diffusivity but as an effective parameter with dimensions of a
diffusivity, related to the capillary coefficient by 
$D = \sqrt{4\epsilon/(1-\kappa)}$. In this interpretation, the SNS
equation provides a quantum wave representation of a purely classical
capillary fluid, in which $\hbar$ and $m$ are a convenient parametrization of $\epsilon$. This is not fully surprising: all classical friction forces have quantum origins.

\subsection{Spherical bubble}

A key advantage of the SNS formulation over the NSK equations emerges
in problems where the density vanishes at the interface, such as the
nucleation of a spherical bubble. Consider a bubble of radius $R$
embedded in a fluid of bulk density $n_0$. We look for a stationary,
spherically symmetric solution $\psi = \Phi(r) \, e^{-i{\bar \mu} t/\hbar}$, with $\Phi(r) =
\sqrt{n(r)}$ and 
$\bar{\mu}$ the chemical potential of the non uniform system. With $U = 0$ and $\gamma = 0$, the SNS equation after Taylor-expanding $\mu(n)$ 
around $n_0$, reduces to:
\beq
\frac{d^2\Phi}{dr^2} + \frac{2}{r}\frac{d\Phi}{dr} =
\frac{1}{\xi_\kappa^2}(\Phi^3/n_0 - \Phi),
\label{eq:bubble}
\eeq
where $\xi_\kappa = \xi \sqrt{1-\kappa}$ is the $\kappa$-dependent
healing length, with $\xi = 
\sqrt{D^2 m/(2 n_0 \mu'(n_0))}$ 
the standard
healing length, 
and ${\bar \mu}=\mu(n_0)$. 
This equation is regular even at $\Phi = 0$, i.e., in the
bubble core where $n \to 0$.

This is a crucial advantage over the NSK formulation. In the NSK
equations, the Korteweg stress contains terms proportional to
$|\bm{\nabla}n|^2/n^2$ and $\nabla^2 n/n$, which are singular when
$n \to 0$. The substitution $\Phi = \sqrt{n}$ in the SNS formulation
removes these singularities, yielding the regular
Eq.~(\ref{eq:bubble}). 
For $R \gg \xi_\kappa$, the curvature term $2\Phi'/r$ is negligible at
the interface $r \simeq R$, and Eq.~(\ref{eq:bubble}) has the approximate solution 
\beq
n(r) \simeq n_0\tanh^2\left(\frac{r-R}{\sqrt{2}\xi_\kappa}\right).
\label{eq:bubble_profile}
\eeq
This approximation predicts a vanishing density on the spherical shell $r=R$. However, treating the curvature term $2\Phi'/r$ perturbatively we find a correction of the density such that we obtain at the interface $r=R$
\beq
n(r)\simeq \left( \sqrt{n_0}\tanh\left(\frac{r-R}{\sqrt{2}\xi_\kappa}\right)-\sqrt{n_0}\frac{\xi_\kappa}{R} \right)^2,
\label{eq:bubble_profile_corrected}
\eeq
which regularized the density and prevents it from reaching zero.
Therefore, the approximate solution Eq.~(\ref{eq:bubble_profile}) is not valid directly at the interface $r=R$ as corrections become relevant. 

The surface tension $\sigma$ of the bubble interface follows from
the profile  of  Eq.~(\ref{eq:bubble_profile}):
\beq
\sigma = 2 \, \epsilon \int
\left(\frac{d\Phi}{dr}\right)^2 dr =
\frac{\sqrt{2}}{6} 
\sqrt{1-\kappa} \frac{D^2 n_0}{\xi},
\label{eq:sigma}
\eeq
which shows that $\sigma$ vanishes for $\kappa \to 1$ (classical NS
limit, no capillary effects) and reaches its maximum for $\kappa = 0$
(Gross-Pitaevskii limit). The Young-Laplace pressure difference across
the bubble interface is as follows:
\beq
\Delta P = \frac{2\sigma}{R},
\eeq
which together with Eq.~(\ref{eq:sigma}) gives a closed expression for
$\Delta P$ in terms of $\kappa$, $D$, $n_0$, $R$ -- all measurable
quantities in a microfluidic experiment.

Finally, it is instructive to compute the Rayleigh dissipation function
for the bubble profile. With $\gamma = \gamma_0$ constant and
$\bm{v} = D\bm{\nabla}\theta$, for a stationary solution with
$\theta = \mathrm{const}$ one has $\bm{v} = \bm{0}$ and therefore:
\beq
\mathcal{R} = \int d^3r\, m\gamma_0 n|\bm{v}|^2 = 0.
\eeq
This result is physically transparent: a stationary bubble with no
flow has zero viscous dissipation. The Rayleigh function $\mathcal{R}$
vanishes identically because the velocity field is zero, not because
$\gamma_0 = 0$. This confirms that $\mathcal{R}$ correctly captures
 dissipative physics: dissipation requires flow, and a static bubble does not flow.

The situation changes if one considers a bubble that grows or shrinks
radially with velocity $\dot{R}(t)$. For an incompressible fluid with
spherical symmetry, the continuity equation $\bm{\nabla}\cdot\bm{v}=0$
gives a velocity field in the exterior region $r > R$:
\beq
\bm{v} = \dot{R}\frac{R^2}{r^2}\hat{r} \; ,
\eeq
which satisfies the boundary condition $v_r(R) = \dot{R}$ at the
bubble interface and vanishes at large distances. The Rayleigh
dissipation function is then:
\beq
\mathcal{R} \, \simeq \, 4\pi m\gamma_0 n_0 \dot{R}^2 R^4
\int_R^{\infty} \frac{dr}{r^2} = 4\pi m\gamma_0 n_0 \dot{R}^2 R^3 \; ,
\eeq
which is finite and proportional to $R^3\dot{R}^2$. The dissipation therefore increases with the bubble volume and the radial velocity, consistently with the expected phenomenology of dissipative interface dynamics.

\section{Sound Modes with Generic $\kappa$ and $\gamma$}

Working with $U=0$ we consider small perturbations around
a uniform stationary state
$\psi_0 = \sqrt{n_0}\,e^{-i\mu(n_0)t/\hbar}$:
\beq
n = n_0 + \delta n, \qquad \theta = \delta\theta,
\eeq
with $\delta n$ and $\delta\theta$ small. The linearized continuity
and momentum equations are as follows:
\beq
\partial_t \delta n + n_0 D\nabla^2\delta\theta = 0,
\eeq
\beq
\partial_t\delta\theta 
+ \frac{\mu'(n_0)}{m D}\delta n 
- (1-\kappa)\frac{D}{4n_0}\nabla^2\delta n
- \gamma D\nabla^2\delta\theta = 0,
\eeq
where $\mu'(n_0) = d\mu/dn|_{n_0}$. Note the sign of the quantum pressure
term: it contributes positively to the restoring force, stabilizing
the system at short wavelengths.

Substituting plane wave solutions
$\delta n,\,\delta\theta \sim e^{i(\bm{k}\cdot\bm{r}-\omega t)}$
and requiring non-trivial solutions, one obtains:
\beq
\omega^2 + i\gamma Dk^2\omega - c_s^2 k^2
- (1-\kappa)\frac{D^2k^4}{4} = 0, 
\label{eq:dispersion}
\eeq
where
\beq
c_s = \sqrt{\frac{n_0 \mu'(n_0) D}{\hbar}}
\eeq
is the speed of sound. Three limiting cases confirm the correctness
of this result. 
When $\kappa = 1$ and $\gamma = 0$, Eq.~(\ref{eq:dispersion}) gives
$\omega = \pm c_s k$ --- linear phonons, as expected for a purely
classical fluid without capillary effects. 
When $\kappa = 0$ and $\gamma = 0$, one obtains:
\beq
\omega^2 = c_s^2 k^2 + \frac{D^2k^4}{4},
\eeq
which is precisely the Bogoliubov dispersion relation of a
weakly interacting Bose gas. The crossover from the phonon regime
($\omega \simeq c_s k$) to the free-particle regime
($\omega \simeq Dk^2/2$) occurs at $k_* = 2c_s/D = \sqrt{2}/\xi$. 
When $\kappa = 1$ and $\gamma \neq 0$, one obtains attenuated phonons.
In the long-wavelength limit $\gamma Dk \ll c_s$:
\beq
\omega \simeq \pm c_s k - \frac{i\gamma Dk^2}{2},
\eeq
with spatial attenuation length $\ell = 2c_s^3/(\gamma D\omega^2)$,
growing as $k^{-2}$ --- consistent with viscous damping in classical
fluids.

For generic $\gamma$ and $0<\kappa<1$, the solution of
Eq.~(\ref{eq:dispersion}) is:
\beq
\omega = -\frac{i\gamma Dk^2}{2} \pm
\sqrt{c_s^2k^2 + \frac{(1-\kappa-\gamma^2)D^2k^4}{4}}.
\label{eq:solution}
\eeq
The quantum pressure $(1-\kappa)$ contributes positively to
$\omega^2$ -- it is a stabilizing dispersive effect that stiffens
the fluid at short wavelengths, consistent with capillary
stabilization of interfaces. The dissipation $\gamma^2$ contributes
negatively -- it tends to suppress short-wavelength
oscillations. The crossover from propagating to purely diffusive
behavior occurs when $(1-\kappa-\gamma^2)D^2k^4/4 > c_s^2k^2$,
i.e., for:
\beq
k > k_c = \frac{2c_s}{D\sqrt{\gamma^2-(1-\kappa)}},
\eeq
which exists only if $\gamma^2 > 1-\kappa$, i.e., when dissipation
dominates over capillary stiffness. Using
Eq.~(\ref{eq:capillary}), this condition reads
$\gamma^2 D^2/4 > \epsilon$, i.e. when viscous dissipation dominates
over capillary effects set by $\epsilon$.

We finally consider the case $\kappa>1$ (and generic $\gamma$), which leads to a negative capillary coefficient $\epsilon$. Under this condition, $(1-\kappa)$ contributes {\it negatively} to $\omega^2$, resulting in a destabilizing effect that weakens the propagation at short wavelengths. Since $\gamma^2$ contributes negatively, both dissipation and dispersion oppose propagation. Thus, the crossover from propagating to diffusive  behavior occurs,  once again, for values larger than $\kappa_c$ which, unlike the previous case, {\it always} exists. This regime may be relevant for building an analog of quantum mechanics to active matter, where negative capillary coefficient (or  surface tension) has been discussed in continuum theories of dry \cite{cates_prx,cates_prl} and wet systems \cite{tiribocchi_prl,singh_prl}.
 
\section{Effective 1D SNS equation for capillary tubes}

A natural application of the SNS equation to microfluidics is the
derivation of an effective one-dimensional equation for a fluid confined
in a narrow capillary tube of radius $a$. This is the fluid analog of
the dimensional reduction from 3D to 1D of the nonlinear
Schr\"{o}dinger equation for Bose-Einstein condensates
\cite{sala2002,lorenzi2024} or electromagnetic waves 
confined in quasi-one-dimensional waveguides \cite{lorenzi2025} 

Consider a cylindrical capillary of radius $a$ and axis along $z$.
We write the wavefunction as:
\beq
\psi(\bm{r},t) = \phi(r)\,\chi(z,t),
\eeq
where $\phi(r)$ is the transverse profile, normalized as
$2\pi\int_0^a \phi^2(r)\,r\,dr = 1$, and $\chi(z,t)$ is the
effective 1D wavefunction. The transverse profile $\phi(r)$ is
determined by the boundary condition at the capillary wall $r = a$
and by the equilibrium condition in the transverse direction.

Substituting into the SNS equation~(\ref{eq:SNS}) with $U = 0$ and
integrating over the transverse degrees of freedom, one obtains the
effective 1D SNS equation:
\beqa
i\hbar\,\partial_t\chi = \left[-\frac{\hbar^2}{2m}\partial_z^2
+ \mu_{1D}(|\chi|^2) + \kappa\frac{\hbar^2}{2m}
\frac{\partial_z^2|\chi|}{|\chi|}\right.\nonumber\\\left.
+ i\gamma_{1D}(|\chi|^2)\frac{\hbar^2}{m}\partial_z^2
\ln\left(\frac{\chi}{|\chi|}\right)\right]\chi,
\label{eq:SNS1D}
\eeqa
where the effective 1D chemical potential is:
\beq
\mu_{1D}(n_{1D}) = \mu(n_{1D}/A_{\rm eff}) + \frac{\hbar^2}{2m}
\langle\phi|-\nabla_\perp^2|\phi\rangle,
\eeq
with $n_{1D} = |\chi|^2$ the 1D number density, $A_{\rm eff} = \pi a^2$
the effective cross-sectional area, and
$\langle\phi|-\nabla_\perp^2|\phi\rangle = 2\pi\int_0^a
|\bm{\nabla}_\perp\phi|^2 r\,dr$ the transverse kinetic energy.
The effective 1D dissipative coefficient is:
\beq
\gamma_{1D}(n_{1D}) = \gamma(n_{1D}/A_{\rm eff}).
\eeq
Eq.~(\ref{eq:SNS1D}) has exactly the same structure as the 3D SNS
equation~(\ref{eq:SNS}), with a renormalized chemical potential and
a dissipative coefficient that depend on the transverse confinement
via $A_{\rm eff}$ and the transverse kinetic energy.

The effective 1D capillary coefficient is the following:
\beq
\epsilon_{1D} = (1-\kappa)\frac{D^2}{4},
\eeq
which is independent of $a$ --- the capillary coefficient along the
tube axis is not renormalized by the transverse confinement.
However, transverse confinement enters through $\mu_{1D}$, which
depends on $a$ through the transverse kinetic energy
$\hbar^2/(2m)\langle\phi|-\nabla_\perp^2|\phi\rangle \sim \hbar^2/(2ma^2)$.

The effective 1D speed of sound is:
\beq
c_{s,1D} = \sqrt{\frac{n_{1D}\,\mu_{1D}'(n_{1D})\,D}{\hbar}} . 
\eeq
 The 1D dispersion relation for sound modes takes the same form
as Eq.~(\ref{eq:dispersion}):
\beq
\omega^2 + i\gamma_{1D} Dk^2\omega - c_{s,1D}^2 k^2
- (1-\kappa)\frac{D^2k^4}{4} = 0,
\eeq
with the crossover wavevector:
\beq
k_c = \frac{2c_{s,1D}}{D\sqrt{\gamma_{1D}^2-(1-\kappa)}}.
\eeq
This shows that the competition between capillary stiffness and
viscous damping in the capillary tube is governed by the same
dimensionless condition $\gamma_{1D}^2 > 1-\kappa$ as in the 3D
case (except for $\kappa>1$), but with the 1D speed of sound $c_{s,1D}$ that depends on
the radius of the tube $a$ through $\mu_{1D}$.

\section{Conclusions and perspectives}

We have shown that the Schr\"odinger-Navier-Stokes equation with generic parameters $\kappa$ and $\gamma$ provides a 
genuine variational formulation of capillary fluid dynamics. 
The action functional Eq.~(\ref{eq:action}) naturally decomposes
 into a conservative part -- containing the Korteweg
capillary stress with capillary coefficient $\epsilon =
(1-\kappa)D^2/4$ -- and a dissipative part given by Eq.~(\ref{eq:rayleigh}). The parameter $\kappa$ thus acquires a precise physical
meaning: it measures the degree of capillarity of the fluid, interpolating
between the Gross-Pitaevskii equation ($\kappa = 0$, maximum capillary
effects) and the classical Navier-Stokes equations ($\kappa = 1$, no
capillary effects). 
The dispersion relation Eq.~(\ref{eq:solution}) shows that $\kappa$
and $\gamma$ play complementary roles: $(1-\kappa)$ controls capillary
stiffness, while $\gamma$ controls viscous damping, with the Bogoliubov
dispersion relation recovered in the quantum limit and viscously damped
phonons in the classical limit. Finally, the dimensional reduction to
an effective one-dimensional equation for capillary tubes opens a
natural avenue for applications to confined microfluidic geometries.

From a computational perspective, it would be interesting to assess
whether the SNS formulation offers advantages over standard
Navier-Stokes solvers, particularly in the presence of capillary
effects, an issue that deserves further investigation.
Moreover, possible connections with analog systems, including 
active or
quantum-inspired fluids, may offer further perspectives \cite{chiofalo_pre}. 
For example, the role of nearcontact ($10$ nm) interactions between macroscopic objects ($100$ micron droplets) 
discussed in \cite{montessori}
is potentially amenable to investigations of quantum macroscopic effects
in soft matter. 
Another example is that effective negative surface tension terms,
obtained with $\epsilon<0$, arise
in active matter systems, where additional non-variational
contributions to the stress tensor may lead to contractile behavior \cite{tiribocchi_prl,singh_prl}. In this context, Eq.(\ref{eq:SNS}) could provide a promising quantum-mechanical analog for describing quorum-sensing bacteria, where rotational contributions are not relevant in the overdamped limit \cite{cates_active}. 
Note also that our theory correctly incorporates both Korteweg tensor and dissipative effects, thus it differs from other formulations, such as that discussed in Ref.\cite{wittkowski_natcomm} where
the dissipation is included via the drag term of the Langevin equation. This suggests that the SNS framework may offer a useful perspective to interpret
such non-equilibrium fluid systems.

It is finally important to emphasize that the search for quantum-like wave formulations of the Navier-Stokes equations has attracted considerable
attention in recent years as a potentially promising route towards
the formulation of quantum algorithms for classical fluid dynamics 
\cite{EPL23,SACHIN25}.
It is estimated that current leading edge computational fluid dynamics simulations on near exa-scale computers, featuring 
$Re \sim 10^{8}$, could be accommodated within $q \sim 80$ qubits, well 
within the nominal capabilities of current quantum hardware.
On the same line, numerical weather forecast on a global scale, featuring 
$Re \sim 10^{13}$ (hence totally unfeasible on a classical computer) could be accommodated within a quantum computer with $q \sim 130$, again
within the current capabilities of current quantum technology (once perfect error correction is in place).
However, in order for such a blue-sky scenario to become real, two major 
obstacles need to be overcome, i.e. non-linearity and dissipation.
Several strategies have been devised to bypass these two barriers, typically
based on Carleman embedding of the nonlinear fluid problem in an infinite-dimensional linear space \cite{SANAVIO24,itani_fluids}. 
However, the resulting quantum algorithms are still far from
achieving quantum advantage.
In this context, it has been argued that casting fluid dynamics 
into a quantum-like wave formalism may help bridge this gap, whence 
a keen interest in SNS formulations \cite{yang2024}. 
By showing that the SNS formulation seamlessly extends to 
non-ideal fluids characterized by complex interface dynamics, we lay down
a potential bridge to the quantum simulation of a broad class of complex
states of flowing matter.

\section*{Acknowledgements}

LS thanks Cesare Vianello and Flavio Toigo for useful discussions. LS acknowledges 
the project ``Frontiere Quantistiche'' (Dipartimenti di Eccellenza) of the Italian Ministry of University and Research
(MUR), the ``Iniziativa Specifica Quantum'' of INFN, the European Union-NextGenerationEU within the National Center for HPC, Big Data and Quantum Computing
(Project No. CN00000013, CN1 Spoke 10: ``Quantum
Computing''), the EU Project PASQuanS 2.  AT acknowledges support from GNFM-INdAM.


\begin{thebibliography}{0}

\bibitem{madelung1926a} E. Madelung, Quantentheorie in hydrodynamischer Form,
Z. Phys. {\bf 40}, 322 (1926).

\bibitem{madelung1926b} E. Madelung, Eine anschauliche Deutung der Gleichung
von Schr\"{o}dinger, Naturwissenschaften {\bf 14}, 1004 (1926).  

\bibitem{vauterin1985} K. Dietrich and D. Vautherin,
  Equivalence between particular types of Navier-Stokes and non-linear
  Schr\"odinger equations, Le Journal de Physique {\bf 46}, 313 (1985).

\bibitem{coste1998} E. Coste, Nonlinear Schr\"odinger equation and
  superfluid hydrodynamics, Eur. Phys. J. B {\bf 1}, 245 (1998). 

\bibitem{sala2008} L. Salasnich and  F. Toigo, Extended Thomas-Fermi
  Density Functional for the Unitary Fermi Gas,
  Phys. Rev. A 78, 053626 (2008).

\bibitem{sala2009} L. Salasnich,  Hydrodynamics of Bose and Fermi superfluids
  at zero temperature: the superfluid nonlinear Schr\"odinger equation,
  Laser Phys. {\bf 19}, 642 (2009). 

\bibitem{molina2016} P. Fern\'andez de C\'ordoba, J. M. Isidro,
  and J. V\'azquez Molina, Schr\"odinger vs Navier-Stokes,
  Entropy {\bf 18}, 34 (2016).

\bibitem{chavanis2017} 
P.H. Chavanis, Derivation of a generalized Schr\"odinger equation from the theory of scale relativity, Eur. Phys. J. Plus {\bf 132}, 286 (2017). 

\bibitem{yang2023} Z. Meng and Y. Yang, 
  Quantum computing of fluidynamics using the hydrodynamic 
  Schr\"odinger equation, Phys. Rev. Research {\bf 5}, 033182 (2023). 

\bibitem{yang2024} Z. Meng and Y. Yang, 
  Quantum spin representation for the Navier-Stokes equation, Phys. Rev. Research {\bf 6}, 043130 (2024). 
  
\bibitem{sala2024} L. Salasnich, S. Succi, and A. Tiribocchi, 
Quantum wave representation of dissipative fluids, 
Int. J. Mod. Phys. C {\bf 35}, 2450100 (2024).

\bibitem{triola2026} C. Triola, Emergence of Vorticity and Viscous
  Stress in Finite Scale Quantum Hydrodynamics, e-preprint arXiv:2508.18200
  to appear in Phys. Rev. E (2026). 

\bibitem{korteweg1901} D. J. Korteweg,
  Sur la forme que prennent les \'equations du mouvements
  des fluides si l'ontient compte des forces capillaires caus\'ees
  par des variations de densit\'e,
Arch. N\'eerl. Sci. Exactes Nat. {\bf 6}, 1 (1901).

\bibitem{dunn1985} J. E. Dunn and J. Serrin,
On the thermomechanics of interstitial working,
Arch. Rational Mech. Anal. {\bf 88}, 95 (1985).

\bibitem{cahn1958} J. W. Cahn and J. E. Hilliard,
Free Energy of a Nonuniform System. I. Interfacial Free Energy,
J. Chem. Phys. {\bf 28}, 258 (1958).

\bibitem{cates_prx} E. Tjhung, C. Nardini, M. E. Cates, Cluster Phases and Bubbly Phase Separation in Active Fluids: Reversal of the Ostwald Process, Phys. Rev. X {\bf 8}, 031080 (2018).

\bibitem{cates_prl} G. Fausti, E. Tjhung, M. E. Cates, C. Nardini, Capillary Interfacial Tension in Active Phase Separation, Phys. Rev. Lett. {\bf 128}, 219901 (2021). 


\bibitem{tiribocchi_prl} A. Tiribocchi, R. Wittkowski, D. Marenduzzo, and M. E. Cates, Active Model H: Scalar Active Matter in a Momentum-Conserving Fluid, Phys. Rev. Lett. {\bf 115}, 118302 (2015)


\bibitem{singh_prl} R. Singh and M. E. Cates,
Hydrodynamically Interrupted Droplet Growth in Scalar Active Matter, Phys. Rev. Lett. {\bf 123}, 148005 (2019).


\bibitem{sala2002} L. Salasnich, A Parola, and L. Reatto, 
  Effective wave equations for the dynamics of cigar-shaped and
  disk-shaped Bose condensates, Phys. Rev. A {\bf 65}, 043614 (2002). 

\bibitem{lorenzi2024} F. Lorenzi and L. Salasnich, 
  Atomic soliton transmission and induced collapse in scattering
  from a narrow barrier, Sci Rep {\bf 14}, 4665 (2024). 

\bibitem{lorenzi2025} F. Lorenzi and L. Salasnich, 
Variational approach to multimode nonlinear optical fibers, 
Nanophotonics {\bf 14}, 805 (2025). 

\bibitem{chiofalo_pre} M. L. Chiofalo, S. Succi,  and M. P. Tosi, Ground state of trapped interacting Bose-Einstein condensates by an explicit imaginary-time algorithm, Phys. Rev. E {\bf 62}, 7438 (2000).

\bibitem{montessori} A Montessori, M Lauricella, N Tirelli, and S Succi, 
Mesoscale modelling of near-contact interactions for complex flowing interfaces, 
J. Fluid Mech. {\bf 872}, 327 (2019).

\bibitem{cates_active} M.  E. Cates, Active Field Theories in Active Matter and Nonequilibrium Statistical Physics, Lecture Notes of the Les Houches Summer School: Volume 112, 2018, edited by Tailleur, J. Gompper, G., Marchetti, M. C., J. Yeomans, M. \& Salomon, C., pp. 180-216 (Oxford University Press, Oxford, 2022)

\bibitem{wittkowski_natcomm} M. te Vrugt, T. Frohoff-H\"ulsmann, E. Heifetz, U. Thiele, R. Wittkowski, From a microscopic inertial active matter model to the Schr\"odinger equation, Nat. Commun. {\bf 14}, 1302 (2023).

\bibitem{EPL23} S. Succi, W. Itani, K. Sreenivasan, and R. Steijl, 
Quantum computing for fluids: Where do we stand? EPL {\bf 144}, 10001 (2023).

\bibitem{SACHIN25} S. S. Bharadwaj and K. R. Sreenivasan,  
Towards simulating fluid flows with quantum computing, Sadhana {\bf 50}, 57 (2025).

\bibitem{SANAVIO24} C. Sanavio and S. Succi. 
Lattice Boltzmann-Carleman quantum algorithm and circuit for fluid flows at 
moderate Reynolds number, AVS Quantum Science, {\bf 6}, 023802 (2024).

\bibitem{itani_fluids} W. Itani and S. Succi, Analysis of Carleman Linearization of Lattice Boltzmann, Fluids {\bf 7}, 24 (2022).


\bibitem{TENNIE} F. Tennie, S. Laizet, S. Lloyd, and L. Magri, Quantum computing for nonlinear differential equations and turbulence, Nature Rev. Phys. {\bf 7}, 220 (2025).

\end{thebibliography}
\end{document}